\documentclass[useAMS,usenatbib,usegraphicx]{mn2e}

\def\kms{km s$^{-1}$}

\def\hii{H\,{\sc ii}}
\def\deg{$^\circ$}
\def\msun{M$_\odot$}
\def\mjyb{mJy beam$^{-1}$}
\def\jyb{Jy beam$^{-1}$}
\def\mdot{M$_\odot$ yr$^{-1}$}
\def\cmtres{cm$^{-3}$}

\def\por{$\times$}
\def\halfa{H$\alpha$}
\def\oiii{[O{\sc iii}]}
\def\sii{[S{\sc ii}]}

\def\h{$^h$}
\def\m{$^m$}
\def\s{$^s$}
\def\aap{A$\&$A}
\def\aj{AJ}
\def\apjs{ApJS}
\def\apj{ApJ}
\def\nar{NewAR}
\def\mnras{MNRAS}
\def\araa{ARA$\&$A}
\def\pasp{PASP}
\def\mjyb{mJy beam$^{-1}$}

\title[Multifrequency study of NGC\,6357. I.]{ A multifrequency study of the active star forming complex NGC\,6357. I. Interstellar structures linked to the open cluster Pis\,24}

\author[C.E. Cappa et al.  ]{ C.E. Cappa$^{1,2}$\thanks{Email:ccappa@fcaglp.fcaglp.unlp.edu.ar}, R. Barb\'a$^{3,4}$, N.U. Duronea$^{1}$, J. Vasquez$^{1,2}$, E.M. Arnal$^{1,2}$, 
\newauthor
W.M. Goss$^{5}$, and E. Fern\'andez Laj\'us$^{2,6}$\\
$^{1}$Instituto Argentino de Radioastronom\'{\i}a (CCT-La Plata, CONICET), C.C. No. 5, 1894 Villa Elisa, Argentina\\
$^{2}$Facultad de Ciencias Astron\'omicas y Geof\'{\i}sicas, Universidad Nacional de La Plata, Paseo del Bosque s/n, 1900 La Plata, Argentina\\  
$^{3}$Instituto de Ciencias Astron\'omicas, de la Tierra y del Espacio (ICATE-CONICET), Av. Espa\~na Sur 1512, J5402DSP San Juan, Argentina \\
$^4$Departamento de F\'{\i}sica, Universidad de La Serena, Cisternas 1200 Norte, La Serena, Chile\\
$^{5}$National Radio Astronomy Observatory, P.O. Box 0, Socorro, NM 87801, USA\\
$^{6}$Instituto de Astrof\'{\i}sica de La Plata (IALP-CONICET), Paseo del Bosque s/n, 1900, La Plata, Argentina}

\begin{document}

\date{Accepted  2000 *****. Received 2010 ****}

\pagerange{\pageref{firstpage}--\pageref{lastpage}} \pubyear{2002}

\maketitle

\label{firstpage}

\begin{abstract}

We investigate the distribution of the gas (ionized, neutral atomic and molecular), and interstellar dust in the complex star forming region NGC\,6357 with the goal of studying the interplay between the massive stars in the open cluster Pis\,24 and the surrounding interstellar matter. 

Our study of the distribution of the ionized gas is based on narrow-band \halfa, \sii, and \oiii\ images obtained with the Curtis-Schmidt Camera at CTIO, Chile, and on radio continuum observations at 1465 MHz taken with the VLA with a synthesized beam of 40\arcsec.  The distribution of the molecular gas is analyzed using $^{12}$CO(1-0) data obtained with the Nanten radiotelescope, Chile (angular resolution = 2\farcm 7). The interstellar dust distribution was studied  using mid-infrared data from the GLIMPSE survey  and far-infrared observations from IRAS. 

NGC\,6357 consists of a  large ionized shell and a number of smaller optical nebulosities. The optical, radio continuum, and near- and mid-IR images delineate the distributions of the ionized gas and interstellar dust in the \hii\ regions and in previously unknown wind blown bubbles linked to the massive stars in Pis\,24 revealing surrounding photodissociation regions. The CO line observations allowed us to identify the molecular counterparts of the ionized structures in the complex and to confirm the presence of photodissociation regions. The action of the WR star HD\,157504 on the surrounding gas was also investigated.  
The molecular mass in the complex is estimated to be (4$\pm$2)$\times$10$^5$ \msun. Mean electron densities derived from the radio data suggest electron densities $>$200 \cmtres,  indicating that NGC\,6357 is a complex formed in a region of high ambient density. The known massive stars in Pis\,24 and  a number of newly inferred massive stars are mainly responsible for the excitation and photodissociation of the parental molecular cloud. 

\end{abstract}


\section{Introduction}

NGC 6357 ($\equiv$ W\,22 $\equiv$ RCW\,131 $\equiv$ Sh2-11) is a large \hii\ complex located in the Sagittarius spiral arm.  The \hii\ complex consists of an incomplete  large shell of about 60\arcmin\ in diameter, many other bright optical nebulosities in different evolutionary stages, many OB stars belonging to the open cluster Pis\,24, and bright infrared (IR) sources, some of them young stellar object (YSO) candidates \citep{lortet84,felli90,bohigas04,wang07,russ10}.

\citet{lortet84} showed that the  large shell was a low-excitation and ionization bounded \hii\ region. This shell, which can also be identified in images in the mid-IR \citep{wang07}, opens to the  north. The shell shows no evidence of expansion motions in optical lines \citep{lortet84} and  has been interpreted as an ionized gas bubble created by the strong winds of the current massive stars in Pis\,24 or by a previous generation of stars \citep{lortet84,bohigas04,wang07}.

\begin{figure*}
\resizebox{17cm}{!}{\includegraphics{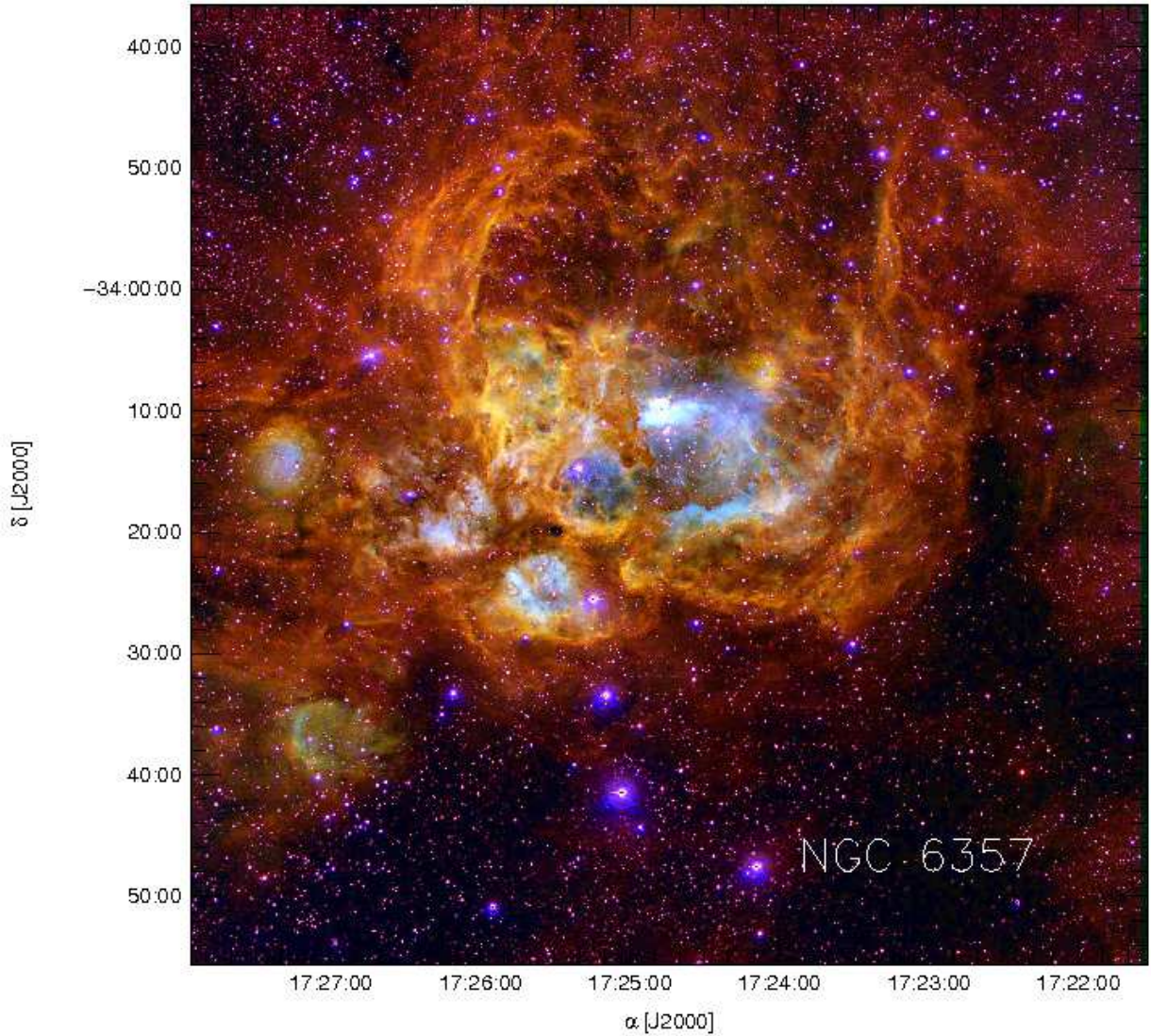}}
\caption{Color composite image of NGC\,6357. Red, green, and blue show emission from \sii\ 6716-31\AA, \halfa, and \oiii\ 5007\AA, respectively. These images were obtained with the Curtis-Schmidt camera at CTIO during 1999.  }
\label{series}
\end{figure*}

The two brightest \hii\ regions in the complex are G353.2+0.9 and G353.1+0.6. G353.2+0.9 is the brightest region at optical and radio wavelengths \citep[e.g.,][]{felli90}. Both extended radio sources were detected at several frequencies \citep[e.g.,][]{shaver70,haynes78}. Detailed studies of these two bright regions were performed using VLA observations at 5 GHz with an angular resolution of 10\farcs 0 \citep{felli90}. Optical images of the whole  NGC 6357  region show a number of additional ionized regions, with no previous optical and radio studies.

The distribution of the molecular gas associated with selected regions in the complex was investigated by   \citet{mcbreen83}  and \citet{massi97}. These last authors found that the bulk of the molecular gas related to the complex has velocities in the range [-14,+4] \kms.

\begin{figure*}
\resizebox{17cm}{!}{\includegraphics{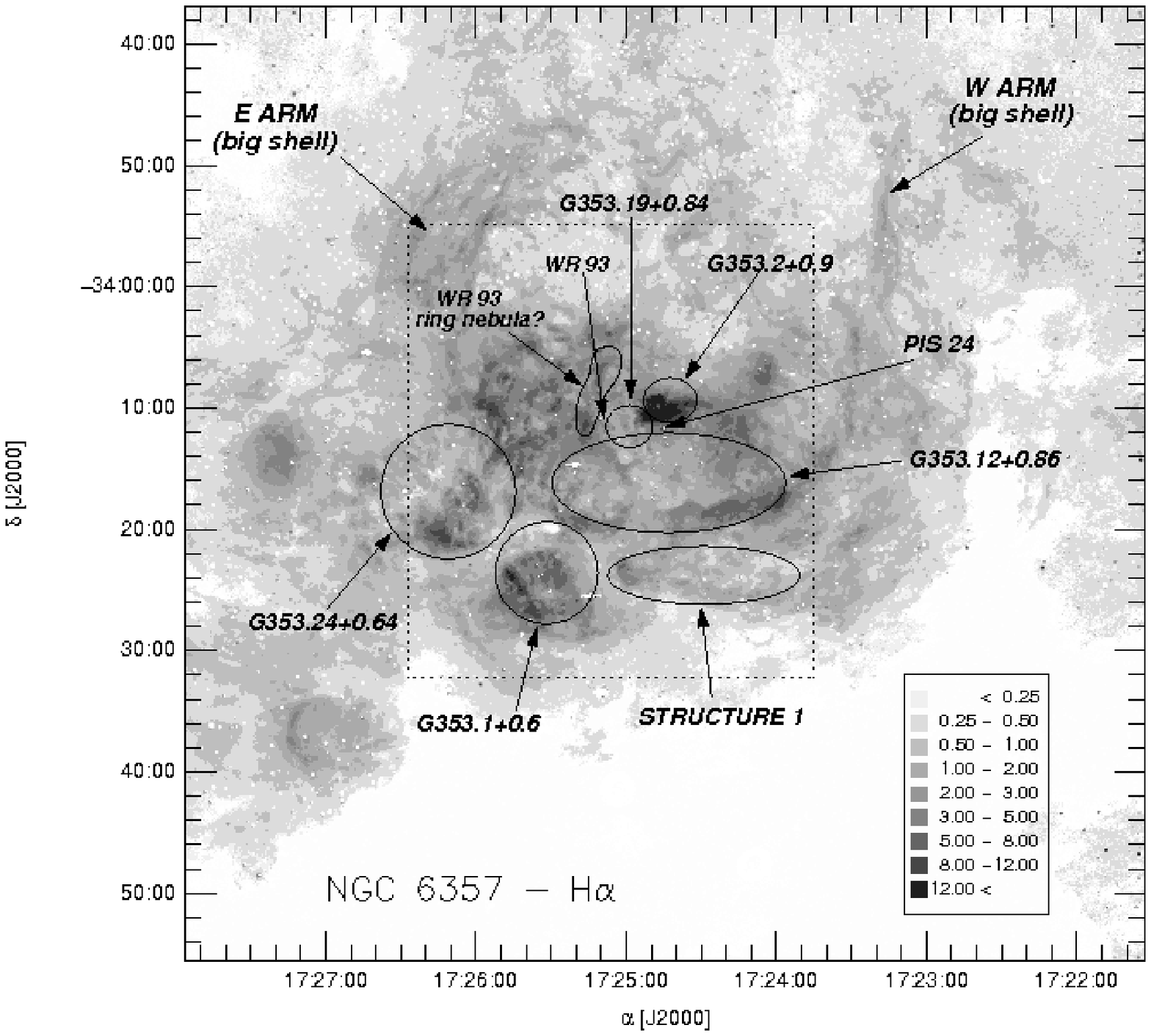}}
\caption{ Continuum-subtracted \halfa\ image of NGC\,6357 obtained at CTIO. Different structures in the ionized complex are indicated to facilitate their identification in Fig. 1. The grayscale values are in units of 10$^{-14}$ erg cm$^{-2}$ s$^{-1}$ pix$^{-1}$. The dotted line delineates the region observed at 1.46 GHz.
 }
\label{series}
\end{figure*}

The massive open cluster Pis\,24, which lies in the central cavity of NGC\,6357 and has an age of about 2\por 10$^6$ yr, is considered to be the only ionization source of the complex because of the huge number of UV photons emitted by the massive stars \citep{massey01}. The cluster contains at least  a dozen O-type stars \citep{wang07}, including two of the most luminous stars known in the Galaxy \citep{bohigas04}. These stars were classified  O3.5If (Pis\,24-1) and O3.5III(f*) (Pis\,24-17) by  \citet{walborn02}. Recent studies by \citet{maizapellaniz07} showed that Pis\,24-1 is a multiple system. The Wolf-Rayet star WR\,93 ($\equiv$ HD\,157406, WC7 + O7-9, \citealt{vanderhucht01}) is also projected onto this region. Massey et al. (2001) consider this star to be a probable member of the cluster. However, the large uncertainties in reddening and distance determinations cast doubts on its relation to the cluster. In addition to the visible stars, massive OB stars hidden in areas with high reddening may also contribute to the ionization and photodissociation of the gas \citep{felli90}.
 
The study by \citet{persi86} of this region focussed on stellar formation activity. This study, along with more recent searches for YSOs in the region, resulted in the discovery of a large number of infrared and X-ray source candidates to YSOs and massive stars with high reddening  \citep[e.g.,][]{bohigas04,wang07}, showing that the whole \hii\ complex is an active area of recent and on-going star formation.

Distance determinations to the ionized complex span the range 1.7 to 2.6 kpc. \citet{neckel78} and \citet{lortet84} derived a distance of 1.7 kpc from UBV and H$\beta$ data, and from the nebular reddening, respectively. \citet{massey01}, based on recent spectral classification and absolute magnitude calibration, redetermined the distance to Pismis 24, adopting  $(m-M)_o$ = 12.0 mag, or $d$ = 2.5 kpc. Following \citet{massey01}, we adopt a distance $d$ = 2.5$\pm$0.5 kpc for the \hii\ complex. 

\begin{figure*}
\resizebox{17cm}{!}{\includegraphics{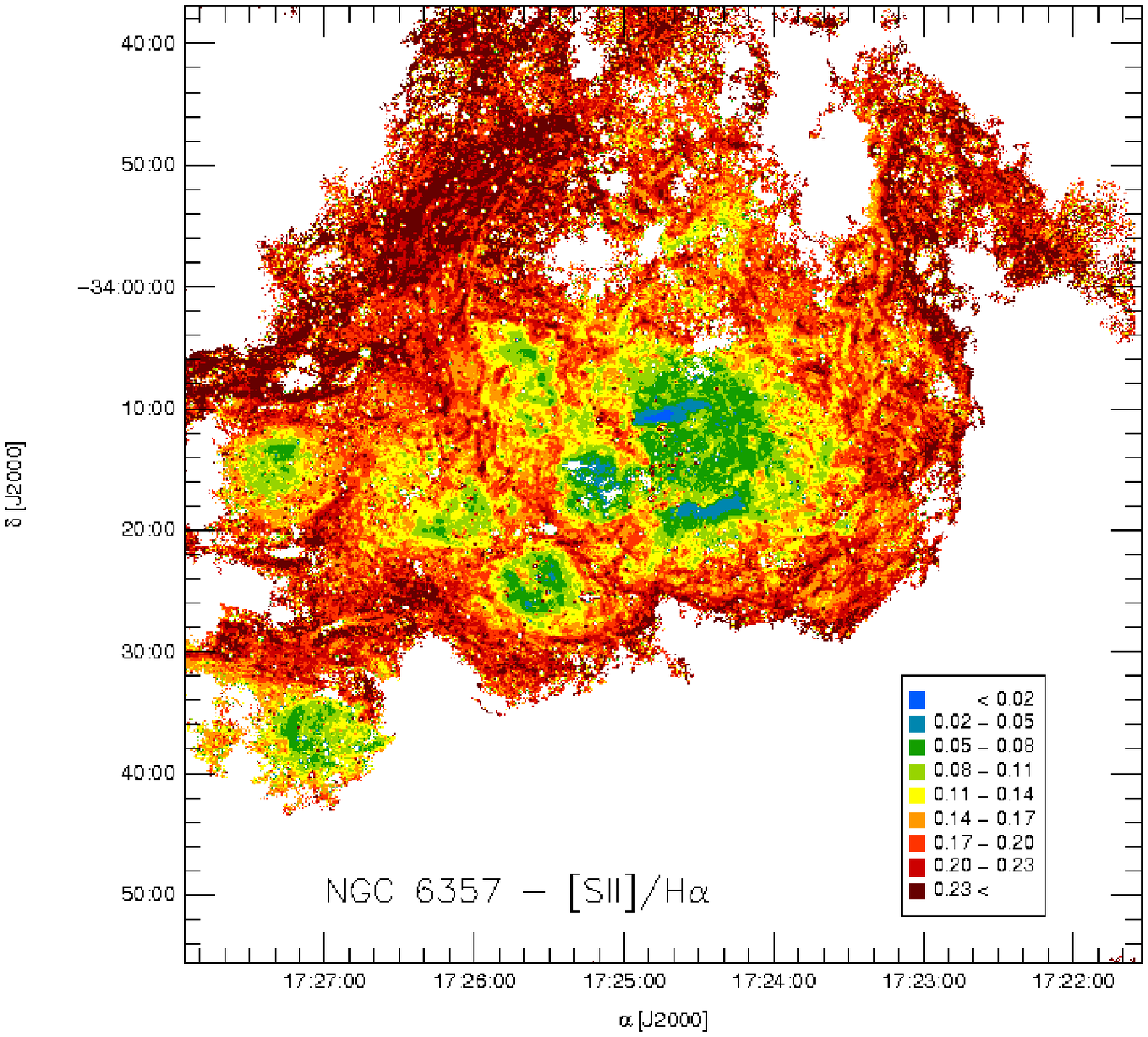}}
\caption{\sii/\halfa\ line ratio using the images obtained at CTIO. Continuum was substracted from the individual images. The color scale is indicated in the lower right corner of the image. White areas correspond to regions below the cutoff value.}
\label{series}
\end{figure*}

   \begin{figure*}
\resizebox{17cm}{!}{\includegraphics{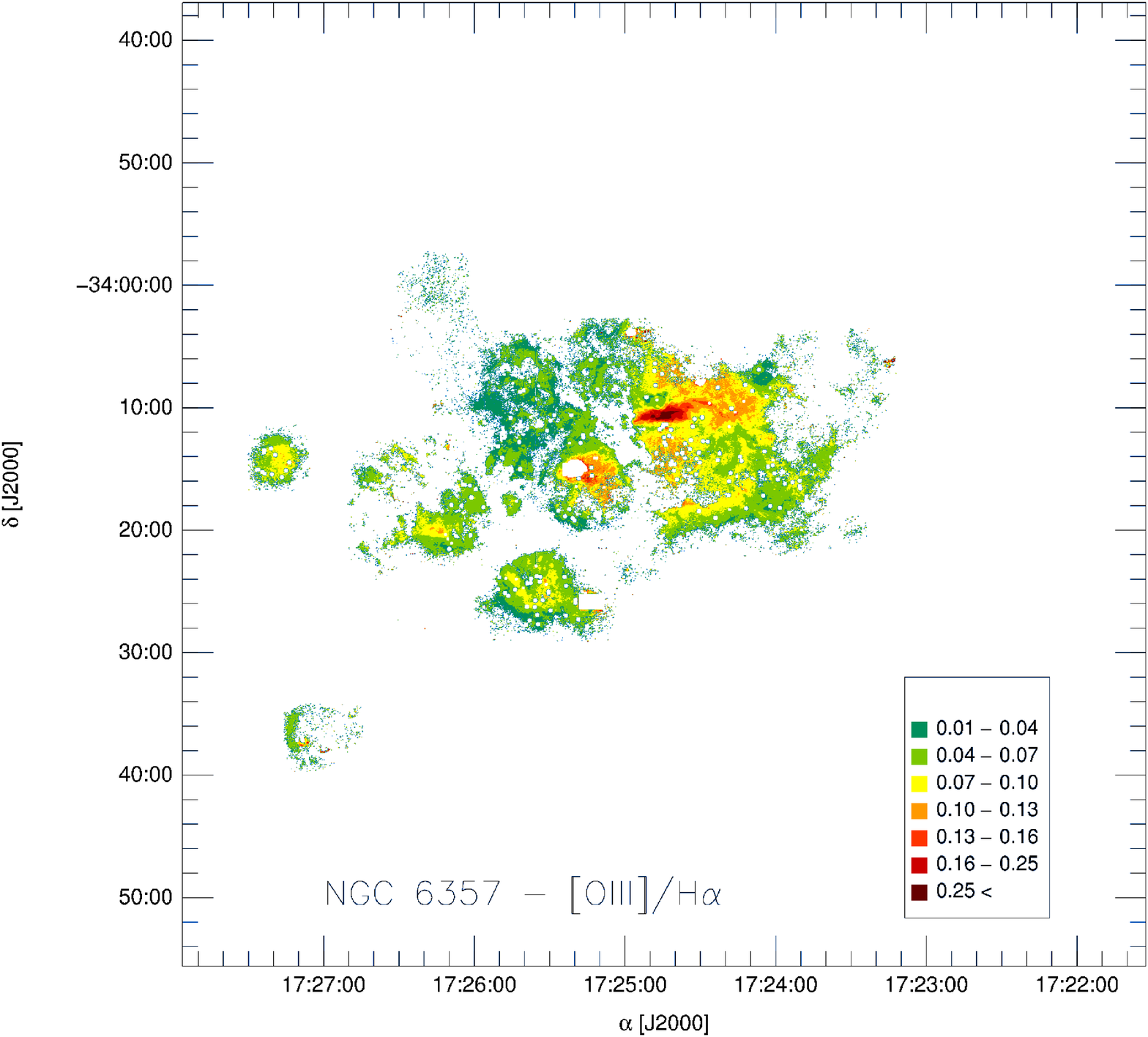}}
\caption{\oiii/\halfa\ line ratio.  Continuum was substracted from the individual images. The color scale is indicated in the lower right corner of the image.   White areas correspond to regions below the cutoff value. }
\label{series2}
\end{figure*}

With the aim of investigating  the interplay between the massive stars belonging to Pis\,24 and the different components of the neighbouring interstellar material, we analyzed \halfa, \oiii\ and \sii\ images obtained at CTIO, VLA radio continuum observations at 1.465 GHz, and CO(1-0) observations obtained using the NANTEN telescope, as well as  IRAS archive images (IRAS-HIRES) at 60 and 100 $\mu$m, and GLIMPSE IRAC archive images in the near- and mid-IR. Optical line ratios are extremely useful in investigating the excitation conditions in different areas of the complex. High resolution radio images lead to the distribution of ionized gas in highly obscured regions, and in the derivation of the physical parameters of  ionized regions. Images in the mid and far infrared lead to characterization of the interstellar dust in the complex. As a substantial amount of molecular gas is expected to be present in star forming regions, we also observed the whole complex in the CO(1-0) line at 115 GHz. 

This is the first of a series of papers dealing with this complex. This paper focusses on some of the most interesting structures in the complex, excited by the massive stars in the open cluster Pis\,24. The bright regions G353.24+0.64 and G353.1+0.6 will be analyzed in a subsequent paper.

Our study provides new information about the interstellar dust, and the ionized and neutral gas distributions in the complex and the dust and gas masses. Interstellar bubbles, \hii\ regions, and photodissociation regions (PDRs) excited by known members of the open cluster Pis\,24 and newly inferred massive stars in the complex are revealed.

\section{Observations}

\subsection{Optical images}

Narrow-band \oiii, \halfa, and \sii, and broad-band $V$ and $R$ CCD images were obtained on May 1999, using the Curtis-Schmidt Camera at Cerro Tololo Inter-American Observatory (CTIO), Chile. The camera was equipped with a Site2K $2048\times2048$ array and has a pixel scale of 2.3 arcsec pixel$^{-1}$. The seeing during observations was typically about 1 arcsec giving a pronounced undersampling for point sources. The narrow-band images were obtained with filters\footnote{Passbands are plotted in\\
 http://www.ctio.noao.edu/instruments/filters/filters$_{34}$.html [www.ctio.noao.edu]} 
centred at 5027\AA, 6567\AA, and 6744\AA, with a FWHM of about 50\AA, 68\AA, and 50\AA, respectively. The \halfa\ filter also includes some contamination of neighbouring [NII] nebular emission lines at 6548\AA\ and 6584\AA.

The individual images in each filter were corrected by bias level and flat-field, and then combined into a single mosaic using IRAF routines\footnote{IRAF is distributed by the National Optical Observatories which is operated by the Association of Universities for Research in Astronomy, Inc., under cooperative agreement with the National Science Foundation.}. These mosaics were registered by using hundreds of stars in the overlaping region. For each final mosaic an astrometric solution was found using stellar positions derived from the Guide Star Catalog 2.0. The absolute coordinate accuracy for each mosaic is better than 0.4 arcsec, although the typical relative uncertainty in the registration between images has been reduced to less than 0.1 pixels.

The surface brightness calibration process was  performed using narrow-band images of Pis\,24 obtained with the Advanced Camera for Surveys (ACS) and the Wide Field and Planetary Camera 2 (WFPC2) on board the Hubble Space Telescope (HST). The datasets correspond to those obtained by Observing Programs No 9857 (PI: O.
De  Marco) and No 9091 (PI: J. Hester) for ACS and WFPC2 instruments, repectively.
The ACS/WFC filter F658N ($\lambda_c \sim6584$, FWHM$\sim$73\AA) has similar characteristics as the \halfa\ filter used in the Curtis-Schmidt camera. The WFPC2 filters F673N ($\lambda_c\sim6732$, FWHM$\sim$65\AA) and F502N ($\lambda_c\sim5012$, FWHM$\sim$37\AA) have also similar characteristics as the \sii\ and \oiii\
filters used in the Curtis-Schmidt Camera. The relative surface brightness zero points were obtained using two areas of about 30 arcsec in common with ACS and WFPC2 images. Uncertainties in the relative calibrations for the  Curtis-Schmidt filters are below 20\%. This figure is derived from the comparison of measurements in adyacent areas.

\subsection{Radio images}

\subsubsection{Radio continuum at 1.46 GHz}

The distribution of the ionized gas was also investigated using radio continuum data. The surveyed field, which corresponds to the central region of the complex,  was observed at 1.465 MHz (20cm) using the Very Large Array in the DnC configuration on 2000 July 3 and 5 as part of the AC555 observing program. The sources 1328+307 ($\equiv$ 3C\,286l, S$_{1.46GHz}$ = 14.9 Jy) and 1748-253 (S$_{1.46GHz}$ = 1.3 Jy) were used as primary and secondary flux density calibrators, respectively. The bandwidth was 50 MHz and the total integration time  2 hours. The coordinates of the field center are 17$^h$25$^m$8\fs 56,--34\deg 11\arcmin 13\farcs 04 (J2000). 
The synthesized beam is 43\farcs 9\por 34\farcs 3 at a position angle of +41\degr. 

The data were edited, calibrated, self-calibrated, and imaged using AIPS 
tasks. The rms in the central part of the image is 20 \mjyb.  

\subsubsection{CO line data}

Intermediate angular resolution and  medium sensitivity {\rm CO} data were obtained with the 4-m {\rm NANTEN} millimetre-wave telescope of Nagoya University. At the time of the observations this telescope was installed at the Las Campanas Observatory, Chile. The half-power beamwidth and the system temperature, including the atmospheric contribution towards the zenith, were  2\farcm 6 and $\sim$220\,K (SSB) at 115 GHz. 

\begin{figure*}
\resizebox{13cm}{!}{\includegraphics{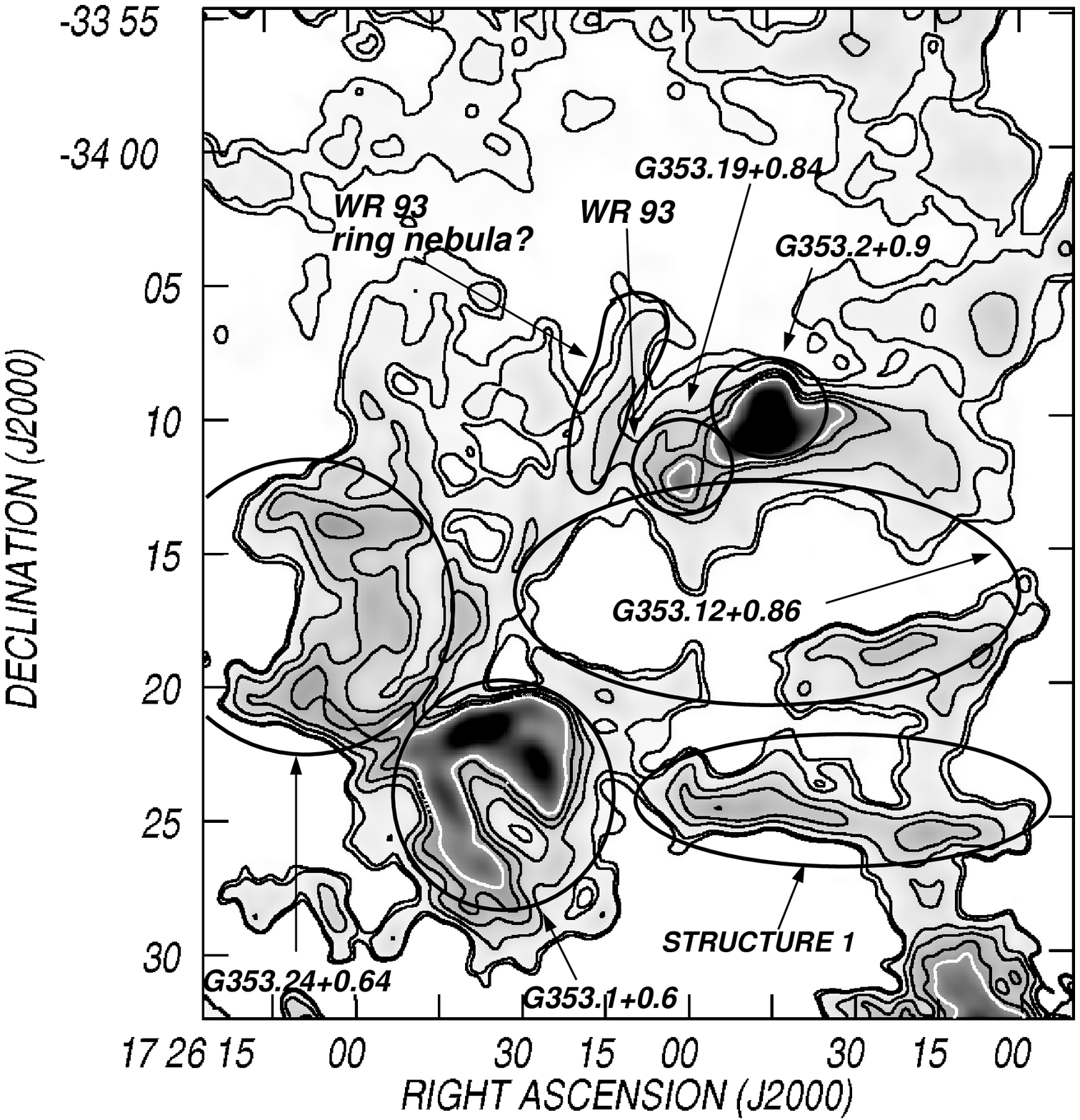}}
\caption{VLA image at 1.46 GHz of the NGC\,6357 complex. Grayscale ranges
from $-$0.01 to 6.6 \jyb. Contours correspond to 0.05, 0.10, 0.30, 0.50, 0.75,
and 1.0 \jyb. Identified structures and the position of WR\,93 are indicated. 
The synthesized beam is 43\farcs 9\por 34\farcs 3 at a position angle of +41\degr. The rms noise in the central part of the image is 20 \mjyb.}
\label{series}
\end{figure*}

The data were gathered using the position switching mode. Observations of points devoid of {\rm CO} emission were interspersed among the program positions. The coordinates of these points were retrieved from a database that was kindly made available to us by the {\rm NANTEN} staff. The spectrometer used was an acusto-optical with 2048 channels providing a velocity resolution of $\sim$ 0.055  \kms. For intensity calibrations, a room-temperature chopper wheel was employed \citet{penzias73}. An absolute  intensity calibration was achieved by observing Orion {\rm KL} (RA,Dec. [J2000] = 5$^h$40$^m$14\fs5, $-$5$^\circ$22\arcmin 49\farcs 3) and $\rho$ Oph East (RA,Dec. [J2000] = 16$^h$32$^m$22\fs8,  --24$^\circ$28\arcmin 33\farcs 1). 
The absolute radiation temperature, T$_R^\ast$, of Orion {\rm KL} and $\rho$ Oph East were assumed to be 65 K and 15 K, respectively  \citep{ulich76,kutner81}.

The {\rm CO} observations were carried out in April 2001 and the surveyed area is defined by 352\fdg51$\leq$~{\it l}~$\leq$353\fdg95 and \hbox{0\fdg14 $\leq$ {\it b} $\leq$ 1\fdg58}\footnote{ In equatorial coordinates, the
four  vertixes of this region are RA,Dec. (J2000) =
(17$^h$25$^m$58\fs21, $-$35$^\circ$10\arcmin 4\farcs 4), 
(17$^h$20$^m$11\fs26, $-$34$^\circ$21\arcmin 9\farcs 8), 
(17$^h$24$^m$6\fs94, $-$33$^\circ$10\arcmin 1\farcs 8), and
(17$^h$29$^m$51\fs35, $-$33$^\circ$58\arcmin 10\farcs 7).}.
An inner area defined by 352\fdg96~$\leq$~{\it l}~$\leq$353\fdg50 and 
0\fdg59$\leq$ {\it b} $\leq$ 1\fdg13\footnote{The
four  vertixes of this region are RA,Dec. (J2000) = (17$^h$25$^m$22\fs9, 
--34$^\circ$32\arcmin 32\farcs 0), 
(17$^h$23$^m$13\fs09, --34$^\circ$14\arcmin 17\farcs 7), 
(17$^h$24$^m$41\fs19, --33$^\circ$47\arcmin 32\farcs 1), 
and (17$^h$26$^m$50\fs64, --34$^\circ$5\arcmin 40\farcs 4).}
was sampled at one beam width intervals, while the remaining area was sampled every 5\farcm4 (two beamwidths). The total number of observed pointings is 349. The integration time per pointing was 16s resulting in a typical rms noise of $\Delta$T$_{rms}$$\sim$0.34 K. 

\subsection{Infrared data}

The {\bf warm} dust distribution was analyzed using high-resolution (HIRES) IRAS data obtained through {\it IPAC}\footnote{{\it IPAC} is funded by NASA as part of the {\it IRAS} extended mission under contract to Jet Propulsion Laboratory (JPL) and California Institute of Technology (Caltech).}. The IR data in the {\it IRAS} bands at 60 and 100 $\mu$m have angular resolutions of 1\farcm 5 and 1\farcm 7. 

High angular resolution IRAC images from the Galactic Legacy Infrared Mid-Plane Survey Extraordinaire (GLIMPSE; \citealt{benjamin03}) at 3.6 $\mu$m (with angular resolution $\phi$  = 1\farcs 5), 4.5 $\mu$m  ($\phi$  = 1\farcs 7), 5.8 $\mu$m ($\phi$  = 1\farcs 8), and 8.0 $\mu$m ($\phi$  = 1\farcs 94) obtained with SPITZER were also used. Additional information on IRAC images is available from \citet{fazio04} and from the Spitzer Science Center Observer Support Website\footnote{Available at http://ssc.spitzer.caltech.edu/ost}.

\section{General characteristics of the complex} 

In this section we  describe the general distribution of the ionized and neutral gas, while a detailed analysis of some of the most interesting regions will be discussed  in the next sections.

Figure 1 shows a composite image of NGC 6357. \halfa\ is shown in green, \oiii\ in blue and \sii\  in red. The filamentary structure of the ionized gas is evident. In addition to the  large shell, clearly detected in \halfa\  and [SII] emissions, the complex shows numerous shell-like features, dust lanes and globules. In Fig. 2 we indicate the location of the most prominent features to facilitate their identification. 
The known \hii\ regions G353.2+0.9, and G353.1+0.6, the  large shell, and a number of other interesting optical structures, like G353.12+0.86 and G353.24+0.64 are indicated in Fig. 2. An elongated structure at RA,Dec.(J2000) =  17\h 24\m 30\s,--34\degr 25\arcmin\ (referred to as Structure 1), an optical filament probably linked to WR 93,   and the positions of WR\,93 and Pis\,24 are also indicated in the figure.

In this paper we investigate the characteristics of the \hii\ region G353.2+0.9 and those of the ionized shell G353.12+0.86 and the  large shell including Structure 1. The prominent regions G353.1+0.6 and G353.24+0.64 will be analyzed in a subsequent paper. 

\begin{figure}
\resizebox{8cm}{!}{\includegraphics{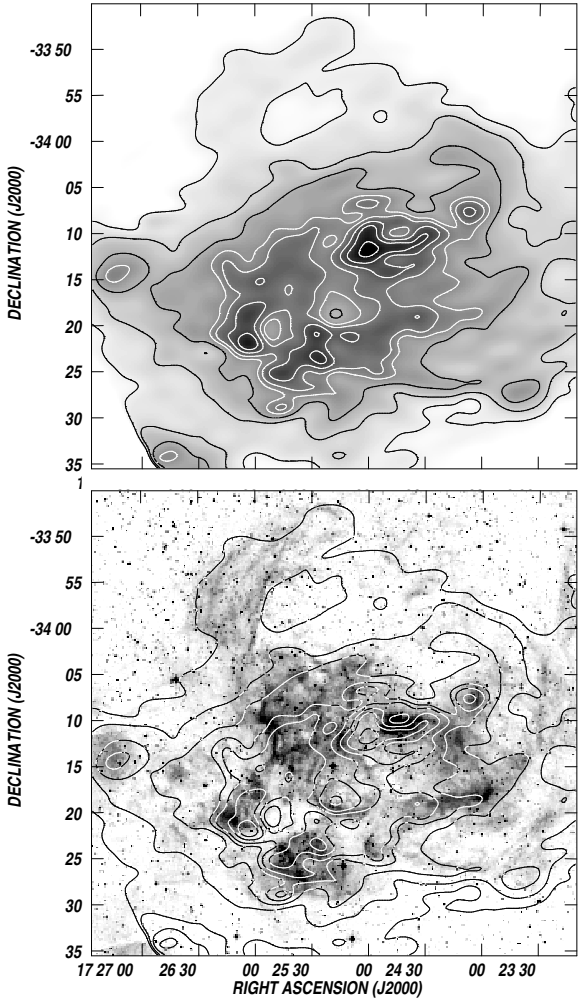}}
\caption{{\it Top panel.} Dust colour temperature $T_c$ estimated from the 
IRAS images at 60 and 100 $\mu$m. Contours range from 33 to 54 K in steps of 3 K. The grayscale ranges from 33 to 54 K (areas in black correspond to the highest temperatures). The image was smoothed to 2\arcmin.
{\it Bottom panel.} Overlay of the dust colour temperature (contours) and 
the H$\alpha$ emission (grayscale). }
\label{IR-1}
\end{figure}

Figures 3 and 4 show the \sii/\halfa\ and \oiii/\halfa\ line ratios. Based on the different excitation conditions, these images provide the evidence to distinguish whether the sources are HII regions, interstellar bubbles or PDRs. Low \sii/\halfa\ and \oiii/\halfa\ line ratios are indicative of low excitation conditions associated with \hii\ regions such as  the  large shell; on the other hand, higher \oiii/\halfa\ line ratios can be associated with interstellar bubbles. In this context, G353.12+0.86 is a stellar wind bubble. 

Figure 5 displays the VLA image at 1.46 GHz, which covers most of the complex. A comparison with the optical images reveals the radio continuum counterparts of most of the optical features indicated in Fig. 2. In particular, the brightest \hii\ regions G353.2+0.9 and G353.1+0.6, which   were studied by \citet{felli90} at 5.0 GHz, are easily identified.  On the contrary, G353.2+0.7 (centered at 17\h 25\m 35.52\s, --34\deg 16\arcmin 54\farcs 6), which was also investigated by \citet{felli90}, can not be identified in the  radio continuum image as a distinct physical structure. The western border of this feature, as delineated by \citet{felli90}, is part of G353.12+0.86, while the eastern section may be part of G353.24+0.64 or of the  large shell, as pointed out by \citet{felli90}. 
The radio counterpart of the shell-like structure G353.12+0.86 is first identified. The radio emitting region at 17\h 25\m, --34\deg 12\arcmin 20\arcsec, located slightly to the southeast of the \hii\ region G353.2+0.9, lacks an optical counterpart.

The upper panel of Fig.~\ref{IR-1} shows the distribution of the dust colour temperature T$_c$ (in contours and grayscale), while an  overlay of the H${\alpha}$ emission distribution  (in grayscale) and  T$_c$ (in contours) is displayed in the lower panel. The dust colour temperature distribution was obtained from the $HIRES$ images at 60 and 100 $\mu$m, following the procedure described by \citet{cichowolski01}. 
Derived temperatures correspond to  $n$ = 1.5.  The parameter $n$ is related to the dust absorption efficiency ($\kappa_\nu\ \propto\ \nu^n$). We adopted $\kappa_\nu$ = 40 cm$^2$ g$^{-1}$, a value derived from the expressions by \citet{hildebrand83}.  Dust temperatures  range from $\sim$25 to $\sim$50 K, with the highest values close to G353.2+0.9, in agreement with previous results from \citet{persi86}. The region of low dust temperatures at 17\h 25\m 15\s, --34\deg 18\arcmin, encircled by regions with higher temperatures, coincides with the eastern section of the cavity of G353.12+0.86. Dust linked to the W arm of the  large shell has a dust temperature of 30 K, while values as high as 50 K were derived for the environs of G353.2+0.9. Regions with high T$_c$ far from Pis\,24 (for example  G353.1+0.6) are suggestive of  the existence of unknown excitation sources {\citep{damke06}}.  

The derived dust temperatures are typical for \hii\ regions. A comparison with the optical image shows that areas with high dust temperature coincide with bright H$\alpha$ or radio continuum emitting regions. This behaviour is compatible with the fact that the stellar UV radiation field of the massive stars in the region is responsible for the heating of the interstellar dust. 

\begin{figure*}
\resizebox{17cm}{!}{\includegraphics{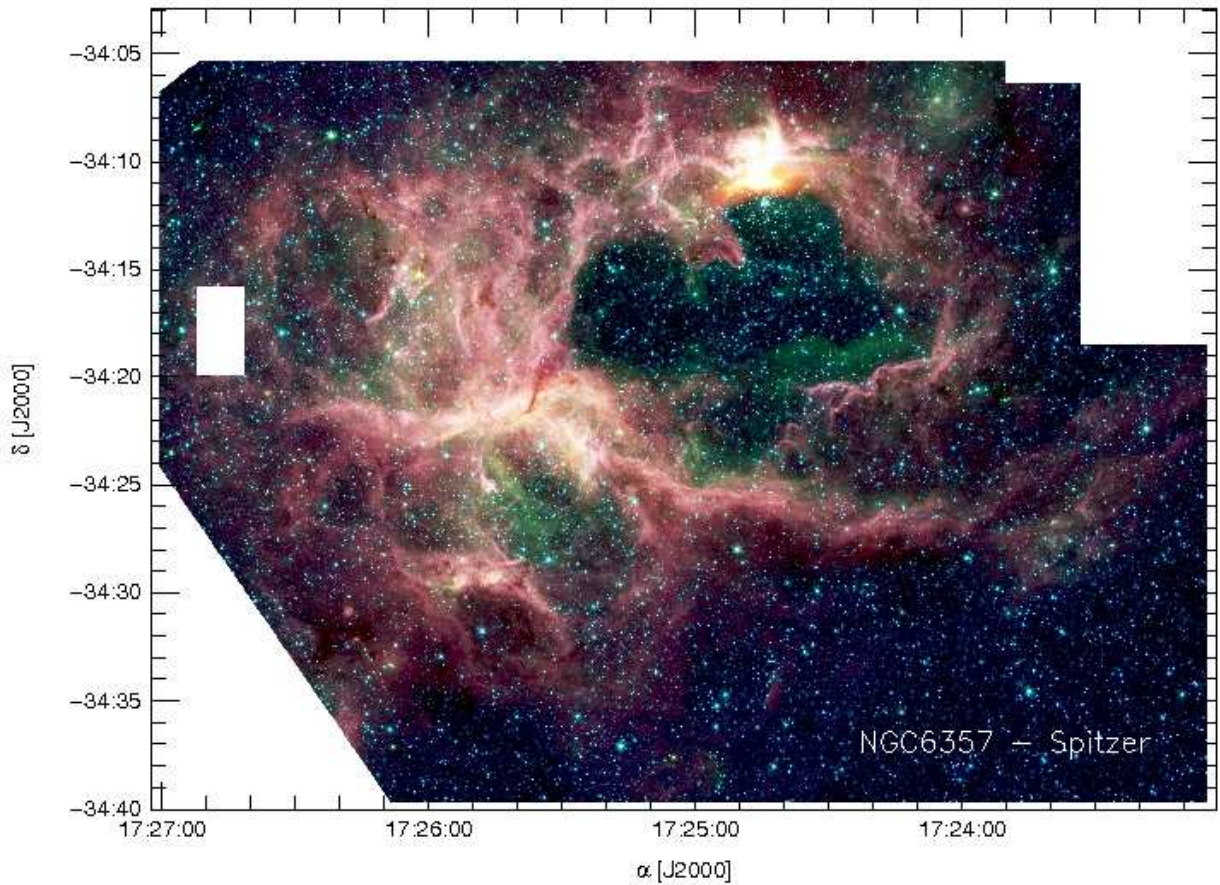}}
\caption{Composite image of the central region of NGC\,6357 using IRAC images: 3.6 $\mu$m is in blue, 4.5 $\mu$m in green, and 8 $\mu$m  in red. }
\label{IR-2}
\end{figure*}

Fig.~\ref{IR-2} displays a composite image of the brightest areas of NGC\,6357 using the IRAC images: 3.6 $\mu$m is in blue, 4.5 $\mu$m in green, and 8 $\mu$m  in red. Emission at 3.6 $\mu$m originates in a faint and diffuse PAH feature at 3.3 $\mu$m and in dispersed stellar light; the 4.5 $\mu$m band shows emission from Br$\alpha$ and Pf$\beta$ and vibrational H$_2$ lines and roto-vibrational CO lines, typical from shocked gas  \citep[e.g.,][]{churchwell06,watson08}. Emission at 8 $\mu$m includes strong features from polycyclic aromatic 
hydrocarbons (PAHs),  which are prominent. 
PAH emission and ionized gas emission (shown by the H$\alpha$ and 1.46 GHz images) have different spatial distributions, with the ionized gas located in the inner area of the structures, closer to the excitation sources than PAH emission. PAHs cannot survive inside \hii\ regions \citep{cesarsky96},  but they do on neutral PDRs at the interface between ionized and molecular gas. Thus, the emission at 8 $\mu$m shows that PDRs are widespread in this complex, a characteristic also indicative of the  presence of molecular gas.
PDRs are clearly delineated in this image (areas in magenta and white). The detection of CI radio recombination lines from several areas in NGC\,6357 is also indicative of the presence of PDRs (Quireza et al. 2006). The emission at 4.5 $\mu$m, coincident with the PDRs (detected at 8 $\mu$m) probably indicates the presence of shocked gas in these last regions. The brightest region in all  IRAC bands is G353.2+0.9. 
 
Elephant trunks are  present throughout the region. Clear examples are located at 17\h 25\m 00\s, --34\deg 14\arcmin\  and at 17\h 24\m 50\s, --34\deg 10\arcmin 50\arcsec\ (this last one first identified by \citealt{bohigas04}). 

Figure 8 displays a series of images showing the \hbox{$^{12}$CO(1-0)} emission distribution within the velocity interval \hbox{[--10.0,+2.5]} \kms\ in steps of 2.5 \kms\ covering the whole  complex. The images show that the bulk of the molecular gas related to the complex has velocities in the range [--10.0,0.0] \kms, coincident with previous findings by \citet{massi97}. 
The most outstanding feature is  a CO depresion near the centre of the images, encircled by a ring-like structure of strong CO emission easily identified in the range [$-$7.5,0.0] \kms (indicated as Shell A in Fig. 8). G353.12+0.86 coincides with the central cavity of Shell A. 

In addition to this structure, other molecular clouds are identified in the CO images. These are labelled clouds B to H and appear to be linked to different nebulosities in the complex. They will be analyzed  in Sect. 4 to 7.   

The large scale CO emission distribution coincides with areas of high extinction and with the massive dense cores identified by \citet{russ10}. 

After this general description, we analyze in the next sections  the  shell  G353.12+0.86, the \hii\ region G353.2+0.9, and the  large shell.

\begin{figure*}
\resizebox{135mm}{!}{\includegraphics{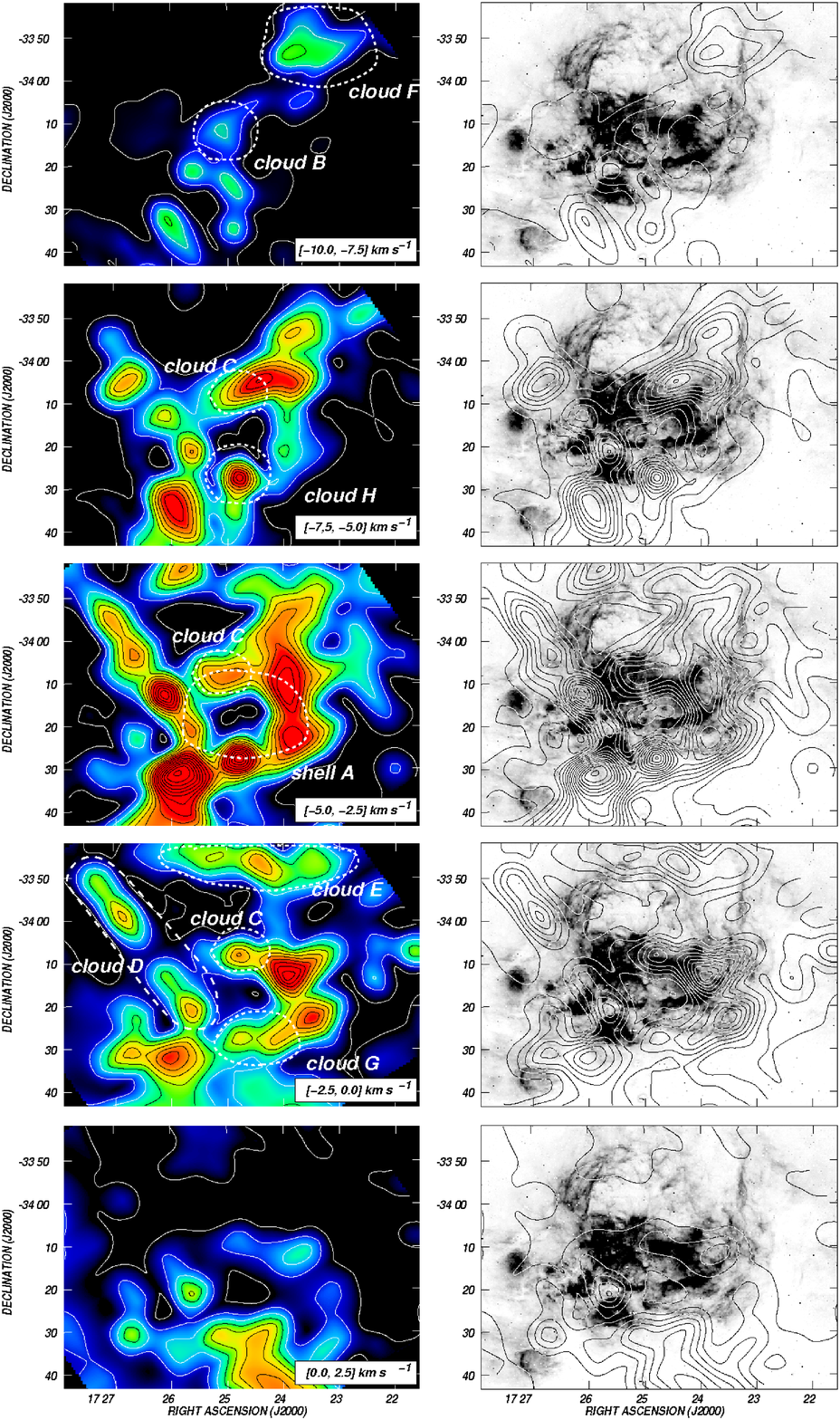}}
\caption{{\bf Figure 8}. {\it Left panels:} $^{12}$CO(1-0) emission distribution within the velocity interval \hbox{[--10.0,+2.5]} \kms\ in steps of 2.5 \kms. The velocity interval is indicated in the bottom right corner of each image. Intensities are expressed as main beam brightness temperature averaged over each velocity interval. 
Contour levels start at 0.7 K ($\equiv$  12.5$\sigma$), increasing in steps of 1.75 K. Colour scale goes from 0.7 (dark blue) to 10.5 K (red). The angular resolution of the CO data is 2\farcm 6. Shell A, as well as clouds B, D, E, F, G, and H are indicated only in one map, although they span a larger velocity interval. 
{\it Right panels:} Overlay of the same CO contours of the left panels and the \halfa image.}

\label{series}
\end{figure*}

\section{The shell-like feature G353.12+0.86}

\subsection{The central region of NGC\,6357}

Two optical shells are apparent near the centre of Fig. 1. They are centred at 17\h 24\m 35\fs, \hbox{--34\degr 14\arcmin 27\arcsec} and at 17\h 25\m 13\s, \hbox{--34\degr 16\arcmin 35\arcsec}, respectively. The central cavities and surrounding envelopes are clearly identifiable in the composite images displayed in Figs. 1, 2, 5, and 7. Figures 1 and 7 reveal that the intense \halfa\ emission that delineates the shells is almost completely encircled by emission at 8 $\mu$m. A comparison of Figs. 1 and 5 shows that both optical shells have radio counterparts.

The optical shells are separated by a dust cloud  clearly discernible in  Fig. 1 \hbox{(at 17\h 24\m 55\s,--34\deg 14\arcmin).}
These images  suggest that this dust cloud is a foreground object. In this scenario, both shells would be a unique structure with the most massive stars  inside.

The CO emission distribution shown in Fig. 8 reveals a region without molecular emission centred at 17\h 24\m 45\s,$-$34\degr 16\arcmin, coincident with the position of the two optical cavities. As pointed out before, the CO hole is almost completely encircled by molecular gas with velocities in the range [--7.5,0.0] \kms\ (indicated as Shell A in Fig. 8). A comparison with the optical image clearly shows that these two optical shells are surrounded by molecular material. Thus, the CO emission distribution confirms the presence of PDRs at the interface between the ionized and molecular gases. 

A comparison of the distribution of the massive dense cores found by \citet{russ10} with Shell A shows that 65\% out of the  $\sim$  70 massive cores are projected onto the CO shell. Velocities of the most massive cores are in the range --2.0 to --4.3 \kms, in agreement with the velocity of Shell A. This indicates that the molecular counterpart of G353.12+0.86 is a region of active star formation  that has probably been triggered by the expansion of G353.12+0.86 (see for example \citealt{hes05}).

The dust column that separates the optical shells can be identified as an elephant trunk pointing towards the centre of G353.12+0.86.
 The dust  cloud coincides in position with a CO cloud having velocities in the range [$-$12.5,$-$7.5] \kms\ (named cloud B in Fig. 8). However, the existence of molecular gas linked to this dust cloud with velocities \hbox{v $>$ --7.5} \kms\ can not be ruled out because of the  poor angular resolution of the CO data in comparison with the optical and radio images. Our results are compatible with those by Massi et al. (1997), who found molecular gas linked to this dust cloud  with velocities in the range [$-$11.5,$-$6.5] \kms\ (the South-Eastern Complex, S.E.C). At least four high density molecular cores listed by \citet{russ10} are projected onto this dust column. For the two most massive cores (\#112 and \#115 in their table 2) these authors find velocities of --5.19 and \hbox{--8.43} \kms, compatible with the velocity of cloud B.    
The fact that the borders of the dust cloud are ionized, the presence of PAH emission at 8 $\mu$m adjacent to the ionized gas {\bf (see Fig.  7)}, and the existence of molecular gas in the region, clearly indicates that molecular gas is beeing photodissociated by the UV photons of Pis\,24.

 As pointed out by \citet{massi97}, the fact that  the molecular gas linked to the dust cloud has velocities more negative than the CO associated with Shell A  gives  additional support to the suggestion that the dust cloud (and Cloud B)  is in front of the shells. Consequently, both ionized shells may be one structure ionized by the UV photons of some of the massive stars in Pis\,24 and swept-up by the strong stellar winds. From here on, this structure of 20\arcmin $\times$9\arcmin\ in size (14.5$\times$6.6 pc at 2.5 kpc), will be referred to as G353.12+0.86, being Shell A its molecular counterpart.

The  molecular hydrogen column density $N_{H2}$ and the molecular mass associated with  Shell A and cloud B  were estimated from the $^{12}$CO data, making use of the empirical relation between the integrated emission $W_{CO}$ ($\equiv \ \int T dv)$ and $N_{H2}$. We adopted $N_{H2}$ = (1.9$\pm$0.3) \por W$_{CO}$ $\times$ 10$^{20}$ cm$^{-2}$ (K \kms)$^{-1}$, obtained by \citet{murphy91}. 
The amount of molecular gas in Shell A between $-$7.5 and 0 \kms\ is  (1.2 $\pm$ 0.6) $\times$10$^5$ \msun, while for Cloud B, we derive (3.4 $\pm$ 1.7) $\times$10$^3$ \msun\  taking into account the emission in the range [$-$12.5,$-$7.5] \kms. Velocity intervals and H$_2$ masses of the clouds identified in NGC\,6357 are listed in Table 2.  

In the following paragraphs we analyze in some detail the characteristics of this structure.  

\subsection{Morphology and characteristics of G353.1+0.86}

The central cavity of G353.12+0.86 
shows diffuse \oiii\ emission revealing that it is filled by hot gas at T $\approx$ 10$^4$ K.
The  \oiii/\halfa\ ratio derived for the cavity (in the range $\approx$ 0.09-0.12, Fig. 4) indicates a region with high excitation conditions. Pis\,24 is the main excitation source of this shell, since most of its massive stars are seen in projection onto the northern part of the cavity.

The northwestern section of the cavity  is sharply bounded by the bright \hii\ region G353.2+0.9 (this region is analyzed in Sect.5).

The western boundary of G353.12+0.86 (near 17\h 24\m 16\s, --34\degr 11\arcmin 38\arcsec) is thick, with an intricate net of filaments detected in optical lines. The \oiii\ filaments are located in the inner part of the wall (Fig. 1). The radio continuum image (Fig. 5) displays  faint and diffuse emission coincident in position with the area where the \oiii\ and \halfa\ filaments are distributed. 
Emission at 8 $\mu$m reveals a complex network of filaments, mixed with those observed at \halfa\  (Figs. 1 and 7). The  emission distribution at different wavelengths in this region suggests that PDRs and optical filaments at different distances along the western wall of the shell are observed.
  
The southweastern rim of the cavity, near 17\h 24\m 30\s,$-$34\degr 19\arcmin, is sharp and bright in optical lines, in the radio continuum, and  at 8 $\mu$m  (Figs. 1, 5, and 7). A close inspection of Fig. 1 shows that the emission in the different optical bands does not coincide in position, being the bright \oiii\ emission closer to the excitation sources than the \halfa\ emission, while the  strong \sii\ emission is located slightly to the south of the \halfa\ emission. The \halfa\ and the radio continuum emissions are closely coincident. The emission distribution at different wavelengths indicates the existence of an excitation gradient, with the high excitation regions closer to the massive stars in Pis\,24 than the low excitation ones. 
The emission at 8 $\mu$m reveals a filament lying $\approx$1\farcm 2 south of the radio emission, indicating that a PDR has developed at the surface of Shell A. 
Emission at 3.6 and 4.5  $\mu$m is also present in these regions (Fig. 9), with the emission at 4.5  $\mu$m almost coincident with the \halfa\ emission. The location of the \sii\ emission farther away from the excitation sources than the \halfa\ emission is typical in PDRs. A similar stratified ionization structure is present at the interface between the ionized gas and the molecular material in the pillars of M\,16, where PDRs have developed \citep{hester96}.

East of 17\h 25\m, the shell displays strong \sii\ emission encircling the H$\alpha$ emitting region. No OB stars (see \citealt{massey01}) are known to be projected onto this section of the cavity.  Inspection of table 5 by \citet{wang07} resulted in the identification of a dozen stars projected onto the cavity. Based on 2MASS colour-colour and colour-magnitude diagrams, these stars are A-type or later.

In summary, the southern, eastern, and western edges of G353.12+0.86 show a clear stratification, with the high excitation ionized gas closer  to the massive stars  in Pis\,24  that the low excitation ionized gas and the PAHs, pinpointing the last ones 
the interface between the ionized and molecular gas.

\begin{figure*}
\resizebox{17cm}{!}{\includegraphics{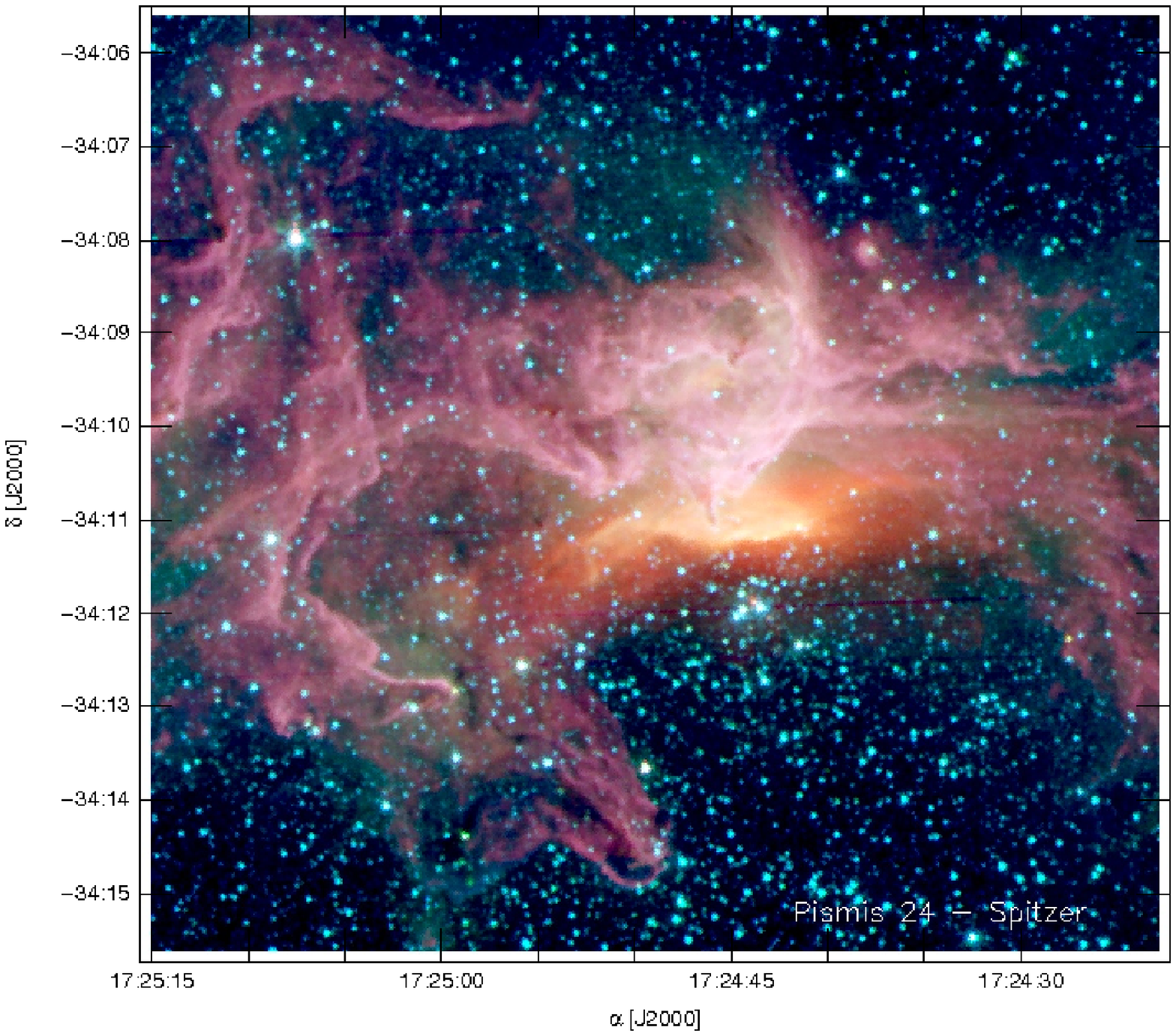}}
\caption{ Enlargement of the G353.2+0.9 region as shown by the emission in the
IRAC bands:  3.6 $\mu$m is in blue, 
4.5 $\mu$m in green, and 8 $\mu$m  in red. }
\label{series}
\end{figure*}

The emission distribution in all bands displayed in Figs. 1, 5 and 7 reveal that the shell is an interstellar bubble blown by the massive stars in the cluster Pis\,24. The interstellar gas has been swept-up and compressed onto the molecular wall by the stellar winds of the massive stars in the stellar cluster. The emission  distribution resembles some of the IR bubbles described by \citet{watson08,watson09}.

The physical parameters of the ionized gas were estimated from the image at 1.46 GHz. These values were derived from the expressions by  \citet{mezger67}. The ionized mass was multiplied by 1.27 to take into account the contribution of He singly ionized (10\% He abundance). We adopted a distance of 2.5 kpc and an electron temperature of 8000 $\pm$ 1000  K, in agreement with estimates from radio recombination lines by \citet{wilson70}  and \citet{quireza06}. We have taken into account that the plasma is distributed in a shell of 1.4 width and covers 20\% of the area of the shell.  Derived values are listed in Table 1, where we include the flux density at 1.46 GHz, the emission measure EM, the size of the structure in arcmin and pc, the rms electron density and  the \hii\ mass and the ionized mass (including singly ionized He).  Errors in electron densities and mases take into account an uncertainty of 20$\%$ in distance are about 50$\%$. 

\section{G353.2+0.9 and its close environs}

Figure 5  reveals that the radio continuum emission in this region is concentrated in two bright clumps: the strongest one coincides with  the optically bright region  G353.2+0.9 (17\h 24\m 47\s, --34\degr 10\arcmin), while the other one  is centered at \hbox{(17\h 25\m, --34\degr 12\arcmin 20\arcsec)} (named G353.19+0.84  from hereon).
A small blister linked to the O6.5V ((f)) star N36 ($\equiv$ Pis 24+3, \citealt{massey01}) was detected by \citet{bohigas04} inside  G353.2+0.9. In the following subsections, we analyze the \hii\ region G353.2+0.9, the small blister linked to N36, and G353.19+0.84.

\subsection{G353.2+0.9}

A comparison between Fig. 1 and 5 suggests that \halfa\ and \oiii\  emissions correlate with the radio emission. The sharp southern boundary detected both at optical lines and in the radio continuum  suggests that the ionized gas is being pushed by the massive stars in Pis\,24, located close to the southern border of
the bright region \citep{bohigas04}. The emission distribution at 8 $\mu$m  shows PAH emission in the IRAC 8 $\mu$m band projected onto the bright \hii\ region  (white area in Fig. 7).

Figure {\bf 9} displays an enlargement of this region in the IRAC bands at  3.6 (in blue), 4.5 (in green), and 8 $\mu$m (in red).  The southern border of G353.2+0.9 is very bright at 4.5 and 8 $\mu$m (area in orange). Emission at 4.5 $\mu$m, detected in  the southern area may originate in shocked gas. The elephant trunk detected at 17\h 24\m 45\s, --34\degr 10\arcmin 55\arcsec\ in optical data by Bohigas et al. (2004), which points directly towards the open cluster, radiates in the PAH features. The southern border of G353.2+0.9 and the pillar were analyzed in detail by \citet{west10}, who found kinematical evidence of strong interaction  between Pis\,24 and the gas in the pillar. Figure 6 shows that the dust colour temperature in this region (near 17\h 25\m 45\s,--34\deg 12\arcmin) is higher than expected for an \hii\ region, which is compatible with a region where shock and ionization fronts are present. 

Figure 8 shows the existence of molecular gas probably related to this \hii\ region with velocities in the range [--7.5,+2.5] \kms. This material partially overlaps the \hii\ region, extending to the north (Cloud C in Fig. 8). CO gas at negative velocities is clearly interacting with the nebula, while material at positive velocities seems to be located behind the ionized gas, in agreement with the detection of PAH emission superimposed to the ionized region.   
These results coincide with those  by \citet{massi97}, who found  molecular gas with peak velocities at --6 \kms\ to the  north and northeast of the nebula, and at --2 \kms\  located behind the ionized gas,  and with \citet{russ10}, who detected dense molecular cores with similar velocities (--3.9 and --5.6 \kms).  
The fact that the ionized region is very bright  in optical lines is compatible with most of the dense molecular gas being to the north or behind the nebula, in agreement with previous findings by \citet{bohigas04} and \citet{wang07}. The molecular gas distribution confirms the previous suggestion by Bohigas et al. that the region is ionization  bounded. 

The amount of molecular gas in the components peaking at $-$6 and $-$2 \kms    (which can not be separated in our data set and correspond to Cloud C), as estimated from our CO data, is  (2.4$\pm$1.2)$\times$10$^4$ \msun.

The physical parameters of the ionized gas, including electron densities and ionized masses derived from our radio continuum image, are summarized in Table 1. We have adopted a background emission of 0.5 \jyb  to perform these estimates, and an electron temperature of 9500 K \citep{bohigas04}. The ionized mass is similar to that estimated by \citet{bohigas04}. Electron densities obtained by those authors from line ratios are about 2000 \cmtres, higher than our estimates (410 \cmtres). An estimate of the filling factor $f$ can be obtained as $f$ = $\sqrt{n_{radio}/n_{opt}}$, where   $n_{radio}$ is the r.m.s. electron density derived from radio data, and $n_{opt}$ is the local electron density estimated from line ratios, which is more sensitive to higher density regions. This parameter indicates the volume of the \hii\ region that is really occupied by plasma. The derived value is $f \approx$ 0.5 and suggests that ionized gas is present in most of the volume of this region.

\subsection{The blister related to N36}

The presence of a small cavity created by the O6.5V ((f)) star N36 (17\h 24\m 45\fs 68,--34\degr 9\arcmin 39\farcs 9, with an optical extintion of about 6.5 mag) inside the brightest section of G353.2+0.9, at 17\h 24\m 46\s, --34\degr 9\arcmin 50\arcsec, is evident in the \oiii\ and \sii\ images, and in the ratio images, mainly in the \oiii/\halfa\ image (Fig. 4).
 
The cavity was already detected by \citet{bohigas04} and \citet{wang07}. The value of the \oiii/\halfa\ ratio changes sharply from about 0.1 in the periphery to about 0.25 in the center of the blister. The \sii/\halfa\ line ratio, which is almost uniform accross the whole region, suggests that the change in the \oiii/\halfa\ ratio is mainly due to  local extinction.

Figure 9 reveals a complex net of filaments  at 3.6 and 8  $\mu$m
(areas in magenta), which also depicts diffuse emission at 4.5  $\mu$m. Most of these concentric
filaments seems to have a common centre, located close to the position of N36.
Small dust regions are also detected in connection with these filaments.

The cavity around N36 can be also identified in the 5GHz VLA image obtained 
with a synthesized beam of 3\farcs 5 by \citet{felli90}. Our VLA image is 
consistent with these findings, since the area of highest radio emission
is projected towards the south of the blister.

\subsection{G353.19+0.84}

The emission at 1.46 GHz extends to the southeast of  G353.2+0.9, where an ionized clump whose brightest section is centered at 17\h 25\m 00\s, --34\degr 12\arcmin 20\arcsec\ can be  identified  (Fig. 5). This clump was barely detected in the image at 5 GHz obtained by \citet{felli90} as a small region of low emission. It can not be identified at  optical lines (Fig. 1). 

This ionized region is behind an area of strong visual absorption described in Sect. 4. The ionized clump is projected onto a ring of PAH emission evident both in Figs. 4 and 9. The emission in the far IR at 60 $\mu$m (not shown here) is relatively strong in all the area. These facts suggest that cold and large interstellar grains are mixed with the ionized gas, where PAHs are destroyed. 

Part of Cloud B, detected in the range \hbox{[--12.5,--7.5]} \kms,  may be either linked to this ionized clump or located  in front of it. 

A striking elephant trunk pointing  towards Pis\,24 appears projected onto the southern border of this radio source (at 17\h 24\m 59.8\s, --34\deg 12\arcmin 50\arcsec). The IR point source located at the top of the pillar (17\h 24\m 59.4\s,--34\deg 12\arcmin 49\farcs 8) might be an evaporating green globule (EGG), as the ones described by \citealt{hester96}. 
   
 The existence  of a PDR bordering the ionized region can not be ruled out.   

The x-ray source at 17\h 24\m 55.85\s, --34\degr 12\arcmin 34\farcs 0 (number 654 in the 
list by  \citealt{wang07}), which has a bright stellar counterpart,  lie
close to the borders of the strong radio source. The MSX sources G353.1998+00.8506 and G353.2021+00.8313, also projected near the borders of the ionized region, can be classified as MYSOs and CHII, respectively, and might be  the excitation sources of
this region. 

The main physical parameters of the ionized gas in this region are summarized in Table 1. A background emission of 0.3 \jyb\ and T$_e$ = 9500 K were adopted.

Finally, parameters for the whole G353.2+0.9 region including G353.19+0.84, and the low level emission region that surrounds these sources are also included in Table 1. An electron temperature of 9500 K was assumed.

\begin{table*}
\centering
\caption[]{Physical parameters of the ionized gas}
\begin{tabular}{lcccccccc}
\hline\hline
      &   $S_{1.46GHz}$ &   $EM$             & 2R         &  2R      & $n_e$    &  $M_{HII}$   & $M_i$      & log \ N$_{Ly-c}$  \\
      &   Jy          &  10$^5$pc cm$^{-6}$ & \arcmin    & pc       & \cmtres  &  M$_{\sun}$  & M$_{\sun}$ &               \\
\hline
G353.12+0.86  & 34.5 & 0.7 & 20$\times$9  & 14.5$\times$6.6 & 200  & 240 & 330 &  50.0       \\ 
\hline
 G353.2+0.9  & 41.0 & 7.8 &  5.4$\times$3.3 & 3.9$\times$2.4 &  410 & 160 & 200 & 49.3    \\ 
 G353.19+0.84& 5.0  & 1.5 & 4.2$\times$2.9 & 3.1$\times$2.1 &   190 &  40 &  50 &     48.5 \\
 Whole area$^1$  & 93.6 & 3.0 & 14.6$\times$5.4 & 10.6$\times$3.9 & 200 &  710 & 900 & 49.7 \\
\hline\hline
\end{tabular}
\vspace{0.3cm}

{\footnotesize  1 Includes G353.2+0.9, G353.19+0.84, and lower emission areas}
\end{table*}

\section{A ring nebula related to WR\,93?}

The bright WR star HD\,157406 (RA,Dec.(J2000) = 17\h 25\m 08\fs 88,--34\degr 11\arcmin 12\farcs 8)   is a probable member of Pis\,24 \citep{massey01}. Spectroscopic studies estimate a mass loss rate of 2.5$\times$10$^{-5}$ \mdot\ and a wind terminal velocity of 2290 \kms\ for the star \citep{prinja90, vanderhucht01}. With these parameters, it is expected that the massive star will deeply perturb the surrounding gas.

The star is projected close to a bright filament seen in optical lines and in the radio continuum at 1.46 GHz, extending from 17\h 25\m 6\s, --34\degr 6\arcmin 25\arcsec\ to 17\h 25\m 15\s, --34\degr 12\arcmin\ (the filament is indicated in Figs. 2 and 5). The filament is particularly bright in \oiii\ and was interpreted as part of a ring nebula related to the WR star  {\citep{marston94}.
  
The emission at 8.3 $\mu$m delineates the eastern border of the filament, reinforcing the suggestion that it is being excited 
by massive stars located to the W of the filament, where the WR star is located. 

The complexity of the gas distribution in NGC\,6357, the large number of shells, filaments and dust patches (in particular near the WR star) insure that the identification of an interstellar bubble created by the strong stellar winds of this star is unlikely. 

\section{The large shell}

The thick E and W arms of the  large shell are clearly identified in the composite image of Fig. 1, with the W arm bending to the east near Dec.(J2000) = --34\degr 27\arcmin.  The ionized filament detected from 17\h 23\m 30\s, --34\degr 27\arcmin\ to approximately 17\h 25\m, --34\degr 24\arcmin\ (named Structure 1 in Figs. 2 and 5) is about 13\farcm 0\por 2\farcm 0 in size. The  large shell, of about 60\arcmin\ in size (or 44 pc at 2.5 kpc) opened to the  north, is mainly detected in \halfa\ emission. Most of the filaments are also easily identified in the faint \sii\ emission (e.g. near RA,Dec.(J2000) = 17\h 25\m 56\s ,--33\degr 25\arcmin 10\arcsec). 
The \sii\ emission is an order of magnitude lower than the \halfa\ emission.  
The high \sii/\halfa\ line ratios ($>$0.17, Fig. 4) of the  large shell and the lack of \oiii\ emission confirm the low excitation conditions, as previously suggested by {\citet{lortet84}}. 

A comparison of the CO emission distribution at different velocities (Fig. 6)
with the \halfa\ emission shows molecular material probably related to the 
  large shell to the east, north and south of the nebula. 
The E arm of the   large shell appears bounded by CO emission located near 17\h 26\m 40\s, --34\degr 5\arcmin\ with velocities in the range \hbox{[--7.5,0]} \kms\
(Cloud D in Fig. 8). This cloud, of about 25\arcmin\ in length, consists of at least three bright clumps and is projected onto a faint \halfa\ emission region. Many dense cores are projected onto two of these clumps \citep{russ10}. We note that part of the material of the CO clumps may belong to Shell A. The coincidence of the molecular emission region  with a region of low optical emission sugests that the molecular gas and the associated interstellar dust are in front of the optical filaments,  in agreement with the existence of  high extinction regions \citep{russ10}. 
 Due to the relatively small field of view, our radio continuum image does not include the E arm and, consequently, it is not clear if optical emission related to this arm is present behind Cloud D.

Towards the north, strong CO emission is also present between --5 to 0 \kms, at --33\degr 45\arcmin, extending from 17\h 23\m\ to 17\h 26\m 30\s\ (Cloud E in Fig. 8), and near 17\h 23\m,--33\degr 56\arcmin\ in the range [--12.5,--7.5] \kms\ (Cloud F in Fig. 8). This material is placed to the  north of the optical filaments of the nebula. Cloud F probably make expansion of the ionized gas towards the  north difficult. 

Structure 1 is detected at 1.46 GHz (Fig. 5) and in \halfa\ and \sii\ lines, being brighter the easthern extreme (at 17\h 25\m 0\fs 5, --34\degr 24\arcmin). The lack of \oiii\ emission is compatible with low excitation conditions. Emission in the IRAC band at 8 $\mu$m borders the southern part of Structure 1 (Fig. 7), extending to the east up to 17\h 25\m 25\s, --34\degr 23\arcmin 20\arcsec, behind a dust cloud. 
The strong emission at 8 $\mu$m to the south of the ionized gas reveals the existence of a PDR at the interface between the ionized and molecular gas.  Structure 1 is projected onto molecular material detected in the range [--2.5,0] \kms\ (Cloud G in Fig. 8), at 17\h 24\m 30\s, --34\degr 30\arcmin. 
CO velocities coincide with the velocity of the ionized gas ($\approx$--7 \kms), as shown by RRL observations obtained by Quireza et al. (2006) towards this area. 

Strong CO emission with velocities in the range \hbox{[--12.5,--7.5]} \kms is present at  17\h 25\m, --34\degr 25\arcmin\ (Cloud H). Part of this material is most probably connected to G353.1+0.6.

The amount of molecular gas associated with the  large shell was estimated by integrating the CO emission in the velocity interval [--12.5,0] \kms. A molecular mass of (5.7$\pm$2.8)$\times$10$^4$ \msun\ was derived for Cloud D, (4.5$\pm$2.2)$\times$10$^4$ \msun\ for Cloud E, (9.5$\pm$4.7)$\times$10$^3$ \msun\ for Cloud F, (2.6$\pm$1.3)$\times$10$^4$ \msun\ for Cloud G, and (3.6$\pm$1.8)$\times$10$^3$ \msun\ for Cloud H. The total amount of molecular gas connected to the outer shell is 1.4$\times$10$^5$ \msun. 

It is worth mentioning that the angular resolution of the CO data is 7 times  larger than that of the radio image, thus making it difficult the association of CO structures with both optical and radio features.

The origin of the  large shell was discussed by Wang et al. (2007). This shell may have originated in the massive stars of Pis\,24. We can not discard the fact that  the massive progenitor of the WR star have contributed to the shaping of the outer shell (e.g. Wang et al. 2007).

\begin{table}
\centering
\caption[]{Parameters of the molecular gas}
\begin{tabular}{lcc}
\hline\hline
                 & (v$_1$,v$_2$) &  M$_{H2}$  \\
                 &  \kms         &  10$^3$ \msun \\
\hline
Shell A          & --7.5,0.0     & 120$\pm$60  \\
Cloud B          & --12.5,--7.5  & 3.4$\pm$1.7  \\ 
Cloud C          &  --7.5,+2.5   & 24$\pm$12  \\
\hline
{\it Outer shell}\\
Cloud D          &  --7.5,0.0    & 57$\pm$28  \\
Cloud E          &  --5.0,0.0    & 45$\pm$22  \\
Cloud F          &  --12.5,--7.5 & 9.5$\pm$4.2\\
Cloud G          &  --2.5,0   & 9$\pm$4  \\
Cloud H          &  --12.5,--7.5 & 3.6$\pm$1.8  \\
\hline\hline
\end{tabular}
\end{table}

\section{Masses, densities, and excitation sources}

The UV photons necessary to keep the gas of the different regions in the complex ionized can be estimated from the radio continuum emission. These values were derived using N$_{Ly-C }$(10$^{48}$s$^{-1}$) = 3.51{\tiny{$\times$}}10$^{-5}$n$_e^2$(cm$^{-3}$)R$^3$(pc). These results are listed in the last column of Table 1.  
Taking into account that at 25-50\% of the UV photons produced by massive stars are absorbed by interstellar dust in \hii\ regions \citep{inoue}, a photon flux of about (3-8)$\times$10$^{50}$ s$^{-1}$ is necessary to maintain G353.12+0.86, G353.2+0.9, and G353.19+0.84  ionized. Ionized gas linked to other regions in the complex was not taken into account. 

Bearing in mind photon flux estimates by \citet{martins02} and Vacca et al. (1996), the massive stars in Pis\,24 can supply a UV photon flux of  (1.4-3.3)$\times$ 10$^{50}$ s$^{-1}$. Although the massive stars in Pis\,24 are major contributors to the ionization of the nebula, in agreement with \citet{massi97} and \citet{bohigas04}, additional massive stars should be identified in NGC\,6357 to explain the ionization of the gas  in the whole complex.

An estimate of the total molecular hydrogen  mass involved in the complex can be derived by integrating the CO emission in the range [--12.5,+5.0] \kms, within the region displayed in Fig. 9. This value turns out to be (4$\pm$2)$\times$10$^5$  \msun. Our estimate also includes molecular gas linked to G353.1+0.6 and G353.24+0.64}.

\section{Summary}

In this paper we have investigated the distribution of the ionized, neutral gas, and interstellar dust towards NGC\,6357. Our goal was to study the interplay between the massive stars in the open cluster Pis\,24 and the surrounding interstellar matter. 

The distribution of the ionized gas was analyzed using narrow-band \halfa, \sii, and \oiii\ images obtained with the Curtis-Schmidt Camera at CTIO (Chile), and radio continuum observations at 1465 MHz taken with the VLA with a synthesized beam of 40\arcsec.  The distribution of the molecular gas and of the interstellar dust were studied using \hbox{$^{12}$CO(1-0)} data obtained with the Nanten radiotelescope, Chile, and near-and  mid-IR data from the GLIMPSE and IRAS surveys, respectively. 

NGC\,6357 consists of a  large ionized shell and  numerous smaller shell-like features, dust lanes, globules and elephant trunks.  \sii/\halfa\ and \oiii/\halfa\ line ratios provide the evidence to distinguish among \hii\ regions, interstellar bubbles, and PDRs. Thus, this study revealed new interstellar bubbles surrounded by photodissociation regions in the complex. Molecular observations allowed us to identify the molecular counterparts of the ionized structures in the complex and to confirm the presence of photodissociation regions. 

The shell G353.12+0.86 is located near the centre of the complex. It is 15.0\arcmin \por 6.8\arcmin\ pc in size  at the adopted distance of 2.5 kpc. The shell is detected in  \halfa\ and \sii\ emission, as well as in the radio continuum at 1.46 GHz. The \oiii\ emission reveals that it is filled by hot gas at 10$^4$ K. PAH emission surrounds the ionized gas emission, indicating the presence of PDRs. A shell of molecular gas was identified in the CO emission distribution, confirming the presence of PDRs. A number of dense cores coincide with the CO shell, indicating that the shell is an active region of star formation, probably triggered by the expansion of G353.12+0.86.  

The  dust column that appears to separate the optical shell in two independient structures is most probably a foreground object. The  difference in velocity between the molecular gas associated with the dust column (in the range [--12.5,--7.5] \kms) and the molecular gas linked to G353.12+0.86 (in the range [--7.5,0.0] \kms\ strongly reinforces this interpretation. In this scenario, G353.12+0.86 is a unique structure with the  massive stars of the open cluster Pis\,24 inside. The emission distribution in the optical, IR,  and radio bands show that this structure is an interstellar bubble blown by the massive stars of Pis\,24. The  interstellar gas has been swept-up and compressed onto the molecular wall.

G353.2+0.9 is the brightest region at optical, IR,  and radio wavelengths.  
The fact that G353.2+0.9 is very bright both in radio continuum and in optical lines is compatible with the location of most of the dense molecular gas to the north  or behind the  nebula, confirming the previous suggestion by Bohigas et al. (2004) that the region is ionization  bounded. 
Electron densities derived from our radio continuum image show that this is the region with the highest rms electron density in the complex. An estimate of the filling factor suggests that ionized gas occupies most of the volume of this region.

The synthesized beam of the radio continuum image allowed us to detect the ionized clump G353.19+0.84, 3.1\por 2.1 pc in size,  located slightly to the  southeast of G353.2+0.9.  This ionized clump is behind an area of strong visual absorption and can not be detected in optical lines. It can be identified in the far IR emission and is partially projected onto a ring of PAH emission.  These characteristics suggest that large interstellar grains are mixed with the ionizad gas, where PAHs are destroyed. 
Two MSX point sources classified as massive young stellar object and compact \hii\ region might be  the excitation sources of this region. 
A striking elephant trunk pointing  towards Pis\,24 appears projected onto the southern border of this radio source. 

The action of the WR star HD\,157504 on the surrounding gas was also investigated. The star may be linked to a bright filament seen in optical lines and in the radio continuum emission. The complexity of the gas distribution near the position insures that the identification of an interstellar bubble created by the strong stellar winds the WR star is unlikely.

The  large shell, with a diameter of 44 pc at 2.5 kpc, opens to the north. It is detected in \halfa\ and \sii\ emissions. The high \sii/\halfa\ line ratios and the lack of \oiii\ emission confirm the low excitation conditions. Molecular gas having velocities in the range [--12.5,+5.0] \kms\ appears related to the easthern, northern, and southern sections of the shell.  The total amount of molecular gas connected to the  large shell was estimated as 1.4$\times$10$^5$ \msun. The massive progenitor of the WR star have probably contributed to the shaping of the  large shell.

Mean electron densities derived from the radio data suggest electron densities excess 200 \cmtres,  indicating that NGC\,6357 is a complex formed in a region of high ambient density. The total molecular hydrogen  mass involved in the complex is estimated as (4$\pm$2)$\times$10$^5$  \msun. 

Estimates of the UV photon flux emitted by the massive stars of Pis\,24 indicate that they are the main contributors to the ionization of the nebula. However, additional massive stars should be identified in  NGC\,6357 to explain the ionization of the gas.

\section*{Acknowledgments}
We thank the anonymous referee for many helpful comments and suggestions, which helped to improve the presentation of this paper.
C.E.C. acknowledge the kind hospitality during her stays at Universidad de La 
Serena, Chile. We acknowledge Jes\'us Maiz Apellaniz for allow us to use JMAPLOT software. The VLA is operated by the National Radio Astronomy Observatory. The NRAO is a facility of the National Science Foundation operated under a cooperative agreement by the associated Universities, Inc. This project was partially financed by the Consejo Nacional de Investigaciones Cient\'{\i}ficas y T\'ecnicas (CONICET) of Argentina under projects PIP 112-200801-02488 and PIP 112-200801-01299,  Universidad Nacional de La Plata (UNLP) under project 11/G093, and Agencia Nacional de Promoci\'on Cient\'{\i}fica y Tecnol\'ogica (ANPCYT) under project PICT 2007-00902. 

This research has made use of the NASA/IPAC Infrared Science Archive, which is operated by the Jet Propulsion Laboratory, California Institute of Technology, under contract with the National Aeronautics and Space Administration.
The MSX mission is sponsored by the Ballistic Missile Defense
Organization (BMDO).
We acknowledge the use of NASA's SkyView facility
    (http://skyview.gsfc.nasa.gov) located at NASA Goddard
    Space Flight Center.
This research has made use of the SIMBAD database and ALADIN software, operated at CDS,
Strasbourg, France.


\end{document}